# The Changing Locus of Health Data Production and Use: Patient-Generated Health Data, Observations of Daily Living, and Personal Health Information Management.

Enrico Maria Piras

Fondazione Bruno Kessler, Center for Information Technology, Via Sommarive 18, 38123 Trento, Italy, piras@fbk.eu

**Abstract.** Despite the growing attention of researcher, healthcare managers and policy makers, data gathering and information management practices are largely untheorized areas. In this work are presented and discussed some early-stage conceptualizations: Patient-Generated Health Data (PGHD), Observations of Daily Living (ODLs) and Personal Health Information Management (PHIM). As I shall try to demonstrate, these labels are not neutral rather they underpin quite different perspectives with respect to health, patient-doctor relationship, and the status of data.

## 1 Introduction

*"I remember that, when I was a child, only the priest had a thermometer. You had to ask him. And you thought twice before you did it!"*

(General Practitioner, Focus Group)

Until quite recently health data production, not to mention its use, has been restricted to healthcare professionals. Laypeople did not have access to tools to gather data with the exception of thermometers or weight scales. Only in some specific conditions, such as diabetes, patients had the instruments to gather and interpret health data and to manage their condition autonomously. In the last ten years, though, the diffusion of miniaturized and easy-to-use measuring devices and availability of consumer health applications and tools are redefining patients into "health information prosumers" (producer-consumer). With some basic computer skills and some low-cost gadgets, it is relatively easy to produce and share an amount of personal health information unimaginable only few years ago.

Health data gathered by patients offer opportunities and pose challenges to existing healthcare systems. For instance, the envisioned forms of technologically-enhanced patienthood will require patients to acquire new skills and it will likely increase their burden of health data management [1] while provider-centric infrastructures will need to be adapted to include those data [2] not to mention the need to develop solutions to allow different communities of users to use the same data for different purposes [3].

Despite the large interest, though, data gathering and information management practices are largely untheorized areas. The purpose of this work is to present and discuss some early-stage conceptualizations emerging in the academic debate and the policy-makers' discourses. I will specifically focus on three labels: Patient-Generated Health Data (PGHD), Observations of Daily Living (ODLs) and Personal Health Information Management (PHIM). I propose here to look at these labels as lenses through which observe the emergence of "patients as data producers", new actors of care provision and management. As I shall try to demonstrate, these labels are not neutral rather they underpin quite different perspectives with respect to health, patient-doctor relationship, and the status of data.

## 2  PGHD, ODL's, PHIM: three acronyms, three approaches

While there is a bulk of literature that discusses about the role of laypeople in producing and managing health data there is no commonly accepted definition of these social practices and of the data gathered. Three emerging labels are gaining relevance in different discourses Despite some relevant differences, there are some significant commonalities in the three approaches. For instance, they all share the idea that patients' role is changing, that these particular data/information will become increasingly important, and that these changes will affect healthcare delivery.. In the next pages I shall try to illustrate their origins and the implicit conceptualizations. To do so I will use a simple analytical grid to map the assumptions underneath each label.

- What **kind of data** the label refers to?
- Which are the main **motives to gather/manage** that data?
- What is the **role of patients/laypeople** with respect to the data?
- What is the **role of healthcare professionals**?
- Which are the **main issues and concerns** regarding this data?
- **Where do data belong**?

### 2.1  Patient-Generated Health Data

Patient Generated Health Data is a label coined by Office of the National Coordinator for Health Information Technology (ONC), a position within the US Department of Health & Human Services. The most comprehensive document regarding PGHD is a white paper published in 2012 [4] and, since then, the label is widely used in official documents [5, 6. Subsequently, the label has spread and has been used in several academic publications.

PGHD are defined as "*health-related data—including health history, symptoms, biometric data, treatment history, lifestyle choices, and other information—created, recorded, gathered, or inferred by or from patients or their designees to help address a health concern*".[4: p. 2]. This definition is a result of the analysis of informal conversations with clinicians, health informatics researchers, patient advocates, health system leaders, and a health law specialists [ibid]. Definition of "data" is broad and includes measured vital signs, self-reported lifestyle data (e.g. diet, exercise) and quality of life data (e.g. sleep quality, social contacts) and they could be "structured or unstructured, machine-readable or not, numeric, text, image, waveform, etc" [ibid: p.4] However, the scenarios used to illustrate the potential of PGHD always refer to structured data gathered through sensors such as glucometers, blood pressure monitors and inhaler with built-in monitoring capabilities.

As pointed out by the definition quoted above, patients create, record and gather data. While patients could make a personal use of PGHD to measure what matters to them and facilitate a patient-defined life [7], the main benefits are foreseen as complements of provider-directed data [6]. Providers are responsible to review PGHD to assess their quality and relevance and to decide whether discarding, documenting in the medical record (EMR), or sharing in the care team but not documented in the record [4].

PGHD pose different challenges to healthcare system: technical, operational legal, and others (for an overview see [8]). The main operational issues, the ones directly faced by healthcare professionals, are associated to the systematic use of PGHD on providers' workflow and staffing. Despite the perceived benefits providers anticipate that they will need more resources to "activate" patients, provide feedback and review PHGD, and identify patient subgroups would be most active in the use of PGHD.

### 2.2 Observations of Daily Living

Observations of daily living is a vernacular introduced by researchers involved in Project HealthDesign, a US program of Robert Wood Johnson Foundation focused on innovation in Ppersonal Health Records [9]. ODLs were "encountered" in the making of the project and are defined as a type of patient-defined and patient-generated data.

*Observations of daily living (ODLs) are the patterns and realities of daily life that have never before considered to be part of one's health record, such as diet, physical activity, quality and quantity of sleep, pain episodes and mood*[10].

While PGHD are mostly provider-defined and may have little or no meaning to patients, ODLs are patient-defined data, "deeply personal, idiosyncratic sensory and behavioural indicators for the purposes of health monitoring and behaviour modification" [11]. ODLs can be collected automatically through sensors or self-reported information, ranging from "counts of nights of adequate sleep to the time frame between eating broccoli and having a bowel movement" [9: p. 6]. Being patient-defined is the foundational feature of ODLs. As such, ODLs cannot be defined a priori but are rather discovered (e.g. when designing personal health management systems) and the key issue for researchers is not to define which data are ODLs rather understanding what motivates and deters people to collect them [12].

Patients do not merely collect data but also decide the tools, the duration and the objectives of data collection. Patient can interpret autonomously ODLs and share (part of) them with providers. Healthcare professionals can assist patients in selecting what and how to collect ODLs and play an important role in motivating patients in continuing data collection. Despite being patient-defined, patient willingness to collect ODLs depends on doctors' willingness to consider them as valuable support to care practices by reviewing ODLs during clinical encounters. In the patient-doctor relationship, ODLs can help provide a richer picture of patient's daily life and cues relevant for case management. This information should be stored in Personal Health Record system, controlled by patients but connected with providers' EMR's.

The main issues regarding ODLs is the increase burden on patients. Collecting health data, especially when it is not automated, is a time consuming activity and motivation to do it may decrease very fast [12]. Sensors or other forms of semi-automated ways to capture ODLs, such as leveraging on line social media [13] are being considered as solutions to this problem.

### 2.3 Personal Health Information Management

Unlike PGHD and ODLs, Personal Health Information Management does not focus on a specific type of data but rather on some social practices that "[…] support consumers' access, integration, organization, and use of their personal health information" [14]. The label has been mostly used in the academic debate and the increasing attention is mostly due to the success of Personal Health Record systems and it has aimed at supporting designers providing them descriptive and analytical tools [14-15]. Most of the works are exploratory in nature and have the purpose to map a largely unknown territory in which health related activities and daily life are inextricably intermingled [15-16].

PHIM refers to the set of activities laypeople, even when they are not experiencing any sickness, perform to manage health related information and varies from person to person, from time to time. A non-exclusive list of motivations to perform PHIM activities include scheduling and planning, coordinating with relatives and caregivers, decision making, tracking, and communicating with peers and healthcare professionals. To these aims, people collect and use a wide array of data and different tools to gather and share them, from family calendars to annotations on healthcare records [15-16]. Despite the wide definition of PHIM, which encompasses all sort of health related information, many papers focus on patient-defined data and on their emergent and (medically speaking) unconventional use.

Doctors have little role in PHIM activities. Most of these activities are not acknowledged by to doctor, nurses and healthcare and remain invisible to them [15-17]. However, healthcare professionals may implicitly require patients to perform some PHIM activities such as having good care of their medical records, keep them in order and bring them at periodic visits [16]. Inadequate PHIM by patients can lead to disruptions in the patient-doctor relationship.

Supporting these activities requires the design of patient-centred tools, flexible enough to accommodate the diverse scopes and the changing needs of users.

The main issues regarding PHIM are identified in the additional burden it causes to patients [17-19], the fragmentation of information collected from various sources [18] and also the willingness of people to share with providers information they perceive "personal" [20]. Moreover, the highly personal styles and tools to gather data may cause difficulties in sharing it with providers.

## 3 The changing locus of health data production and use. Three perspectives: delegation, self-care, empowerment

Policy-makers, healthcare managers and professional, vendors, medical informatics scholar, and patients are showing growing attention towards data produced and managed outside the healthcare settings. These actors have different expectations about the risks and benefits of this data, about new forms of patienthood and patient-doctor relationship, about novel kind of healthcare provision.

These expectations are often implicit and hidden behind general statements. In this work I've tried to unpack these unspoken imaginaries through the analysis of three emergent labels: Patient-Generated Health Data (PGHD), Observations of Daily Living (ODLs) and Personal Health Information Management (PHIM). All the three labels adopt quite broad, and often overlapping, definitions. The overlapping of definitions is drastically reduced once we scrutinize how each label is used in practice (see Table 1).

**Table 1.** PGHD, ODLs, and PHIM

|  | **Patient-Generated Health Data** | **Observations of Daily Living** | **Personal Health Information Management** |
|---|---|---|---|
| **Kind of information** | Structured, gathered through sensors | Patient-defined (structured and discursive) | All health related information |
| **Motives to gather/manage** | Help address a health concern | Improve health; manage chronic condition | Health management (in the broadest sense) |
| **Role of patients / laypeople** | Collect and share data | Collect, share, classify and interpret data | Collect, share, classify and interpret data |
| **Role of professionals** | Review data | Encourage collection; review data | Make sense of unstructured data |
| **Where data belong** | EMR (if validated) | PHR (connected with providers' systems) | PHR, other personal health management systems |
| **Main issues and concerns** | Burden on health professionals | Burden on patients; little relevance for providers | Burden on patients; difficult to use by providers |

PGHD is used to refer to structured health data patients are delegated to gather by their providers to complement provider-collected information. Provider assess data and decide if it deserves to be included in the medical records. Data are mostly structured and its relevance is evaluated by providers that use it to help address a medical-defined health issue. PGHD, thus, is used with reference to a provider-driven care model characterized by an asymmetric patient-provider relationship.

ODLs refers to patient-defined data that can help patient in self-managing practices. While this data can be shared with providers to have a richer picture of patients' health its collection can be motivated by the desire to improve one's condition autonomously. ODLs can be collected for self-managing purposes and thus require skills to define what to track and how to make sense of the data. ODLs is used to refer to a multifaceted health management model which envisions a more balanced pattern of patient-provider relationships coupled with patient-defined health/wellbeing goal setting.

Finally, PHIM refers to the mundane practices of managing health information in everyday life. Boundaries between health and other spheres of activities blur as practices blend together. PHIM is used to highlight the unfinished business of personal health management and the tinkering to accommodate personal life and providers' recommendations.

The three labels presented are not neutral or interchangeable. Rather, they are part of larger set of envisioned relationships, expectations regarding roles and provision of care in the next future and the associated concerns. Each label represents a lens through which consider the challenges and the possibilities offered by technologies that are changing patients into data prosumers.